\documentclass[fleqn,twoside,twocolumn,nofootinbib,showkeys]{revtex4}
\usepackage[sec,nocpr]{ujp}
\usepackage{graphicx}
\usepackage{hyperref}
\usepackage{epsfig}
\usepackage{epstopdf}
\usepackage{color}

\begin{document}
\title[Higgs boson decay to lepton pair and photon  and possible non-Hermiticity of the Yukawa interaction]%
{Higgs boson decay to lepton pair and photon and possible non-Hermiticity of the Yukawa interaction}%
\author{V.A.~Kovalchuk}%
\affiliation{NSC ``Kharkiv Institute of Physics and Technology'' }
\address{Akademichna, 1, 61108 Kharkiv, Ukraine}
\affiliation{V.N.~Karazin Kharkiv National University}%
\address{Svobody Sq., 4, 61022 Kharkiv, Ukraine}%
\email{koval@kipt.kharkov.ua}
\author{A.Yu.~Korchin}%
\affiliation{NSC ``Kharkiv Institute of Physics and Technology'' }%
\address{Akademichna, 1, 61108 Kharkiv, Ukraine}
\affiliation{V.N.~Karazin Kharkiv National University}%
\address{Svobody Sq., 4, 61022 Kharkiv, Ukraine}%

\razd{\seci}

\autorcol{V.A. Kovalchuk, A.Yu.~Korchin}%

\setcounter{page}{557}%

\begin{abstract}
The production of lepton pairs in the Higgs boson decay $h \to
\ell^+ \ell^- \gamma$ is studied. The emphasis is put on the
structure of the Higgs boson interaction with the fermions. This
interaction is chosen as a mixture of the scalar and pseudo-scalar
couplings, and, in addition, it is supposed to be non-Hermitian.
We study prediction of this model for the observables in the $h
\to \ell^+ \ell^- \gamma$ decay for the $e^+ e^-$, $\mu^+ \mu^-$
and $\tau^+ \tau^-$ pairs. The differential decay width and lepton
forward-backward asymmetry are calculated as functions of the
dilepton invariant mass for several sets of $h f \bar{f}$ coupling
constants. The influence of non-Hermitian $h f \bar{f}$ interaction on the
forward-backward asymmetry is studied,  and large influence of a
possible non-Hermiticity of the Higgs interaction with the top
quarks on forward-backward asymmetry for $e^+ e^-$ and $\mu^+
\mu^-$ pairs is stressed.

\end{abstract}


\maketitle

\section{\label{sec:Introduction}Introduction}

In 2012 the collaborations ATLAS and CMS at the Large Hadron
Collider (LHC) discovered the spin-less particle $h$ with the mass
about 125 GeV~\cite{ATLAS:2012,CMS:2012}. The study of processes
of production of $h$ boson and its decay channels allowed to
conclude~\cite{CMS:2015,ATLAS:2016}, that its characteristics are
consistent with properties of the Higgs boson of the Standard
model (SM). In particular, the analysis of the angular
correlations in the decays $h\to
ZZ^*,\,Z\gamma^*,\,\gamma^*\gamma^*\to 4\ell$, $h\to WW^*\to
\ell\nu\ell\nu$ ($\ell=e,\,\mu$), and $h\to \gamma\gamma$ revealed
that all the data agree with predictions for the Higgs boson with
the quantum numbers $J^{\rm PC}=0^{++}$~\cite{CMS:2012sp,ATLAS:2013,CMS:2015sp}.

The masses of the fermions in the SM arise due to the Yukawa
interaction between the Higgs field and the fermion fields. The
investigation of this interaction is necessary for identification
of the particle $h$ with the Higgs boson of the SM. Particularly,
it is important to check whether the Higgs boson interaction with
fermions is Hermitian or not. Moreover, it is necessary also to
verify the Hermiticity of the Yukawa interaction
Lagrangian~\cite{Korchin:2016}.

Presently for the decay modes $h\to \tau^+\tau^-$ and $h\to b\,
\bar b$ the Higgs signal strength parameter $\mu(X)$ is
determined. This is the ratio of the experimentally measured cross
section of the Higgs boson production with subsequent decay into
certain final state $X$, to the corresponding value calculated in
the SM:
\begin{equation}
\mu (X) = \frac{\sigma(pp \to h)_{exp} {\rm BR}(h \to X)_{exp} }{\sigma(pp \to h)_{\rm SM} {\rm BR}(h \to X)_{\rm SM}}.
\label{eq:000}
\end{equation}

There are ATLAS and CMS measurements of
the production cross section and decay rate of the Higgs boson, as
well as the constraints on the coupling constants of the
interaction with vector bosons and fermions.
The result of this analysis are the values $\mu (\tau^+ \tau^-) = 1.12 \pm 0.23$
and $\mu (b \bar{b}) = 0.82 \pm 0.30$~\cite{PDG:2016}.

Some aspects of possible non-Hermiticity of the Higgs boson
interaction with the top quark have been studied
in~\cite{Korchin:2013a,Korchin:2013b,Korchin:2014}. Namely,
in \cite{Korchin:2013a,Korchin:2013b}, the polarization
characteristics of the photon in the decay processes $h \to \gamma
\gamma$ and $h \to \gamma Z $ were studied. The photon circular polarization in
these decays arises due to presence of the $\mathcal{CP}$-even and
$\mathcal{CP}$-odd components in the $h t \bar{t}$ interaction,
small imaginary loop contribution in the SM, and non-Hermiticity
of the $h t \bar{t}$ interaction. In \cite{Korchin:2014}, it was
shown that the lepton forward-backward asymmetry $A_{FB}$ in the
processes $h \to \gamma \ell^+ \ell^-$ (for $\ell= e, \mu, \tau$)
is sensitive to a possible non-Hermiticity of the Higgs interaction with the top quark.
We emphasize that measurement of any
observable which is sensitive to non-Hermiticity of Lagrangian
can be used at the same time for testing the $\mathcal{CPT}$
theorem, since Hermiticity of Hamiltonian (or Lagrangian) is the
necessary condition in the proof of the $\mathcal{CPT}$ theorem in
the quantum field theory (see, {\it e.g.}, \cite{Streater:1964}).

In the present paper, we consider the influence of a possible
non-Hermiticity of the Higgs boson interaction with fermions
(leptons and quarks), which contains both scalar and pseudoscalar
parts, on the decay $h \to \gamma \ell^+ \ell^-$, where $\ell =
(e, \, \mu, \, \tau)$. We calculate the differential decay rate
and lepton forward-backward asymmetry as functions of the
invariant mass of the lepton pair, and discuss the integrated over
the invariant mass decay rate and the asymmetry.

\section{\label{sec:formalism} Decay amplitudes and angular distribution }

We assume that the interaction of the $h$-boson with the fermion fields,
$\psi_f$, is described by the Lagrangian
\begin{equation}\label{eq:001}
{\cal L}_{hff}=-\sum_{f = \ell, \, q} \frac{m_f}{v}\,h\,{\bar
\psi_f}\left(a_f+i\,b_f \gamma_5\right)\psi_f \,,
\end{equation}
which includes both scalar and pseudoscalar parts.
Here $v=\left(\sqrt{2}G_{\rm F}\right)^{-1/2}\approx 246$ GeV is
the vacuum expectation value of the Higgs filed, $G_F =
1.1663787(6) \times 10^{-5}$ GeV$^{-2}$ is the Fermi
constant~\cite{PDG:2016}, $m_f$ is the fermion mass, and $a_f$,
$b_f$ are complex-valued parameters (of course, $a_f=1$ and
$b_f=0$ correspond to the SM). One can consider~(\ref{eq:001}) as
a phenomenological parameterization of effects of ``new physics'' (NP)
beyond the SM. At the same time, the interaction with the $W^\pm$
and $Z$ bosons is taken as in the SM. For real values of
parameters $a_f$ and $b_f$ the interaction~(\ref{eq:001}) is
Hermitian. However, it is straightforward to
verify~\cite{Korchin:2016} that the width of the $h$-boson decay 
to the fermion pair in lowest order is the same for Hermitian 
and non-Hermitian Lagrangian (\ref{eq:001}).

We consider the $h$-boson decay
\begin{equation}\label{eq:002}
h(p) \, \to \, \gamma (k, \, \epsilon(k) ) +\ell^+ (q_+) + \ell^-
(q_-) \,,
\end{equation}
where the 4-momenta of the $h$ boson, photon and leptons are,
respectively,  $p$, \ $k$, \  $q_+$, \ $q_-$, and $\epsilon (k)$
is the 4-vector of the photon polarization.

The differential decay width can be written as
\begin{equation}
\frac{ {\rm d} \Gamma }{{\rm d}  q^2 \, {\rm d} \cos\theta }=
\frac{\beta_{\ell}  (m_h^2 -q^2)}{(8 \pi)^3 \, m_h^3} \,\sum_{\rm
pol} |{\cal M}|^2 \,, \label{eq:003}
\end{equation}
where $m_h$ is the mass of the $h$ boson, $q \equiv q_+ + q_- $, \
$q^2$ is the invariant mass squared of the lepton pair,
$\beta_{\ell} = \sqrt{1-4 m_{\ell}^2/q^2}$ is the velocity of
lepton in the rest frame of the lepton pair. The polar angle
$\theta$ is determined in this frame; it is the angle between the
momentum of $l^+$ lepton and the axis which is opposite to the
direction of motion of the Higgs boson.

\begin{figure}[tbh]
\begin{center}
\includegraphics[width=0.38\textwidth]{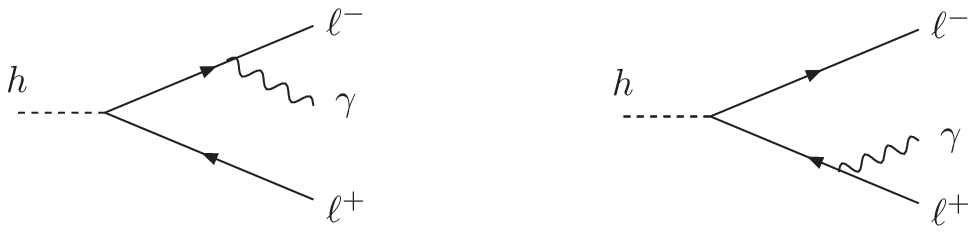}

\includegraphics[width=0.49\textwidth]{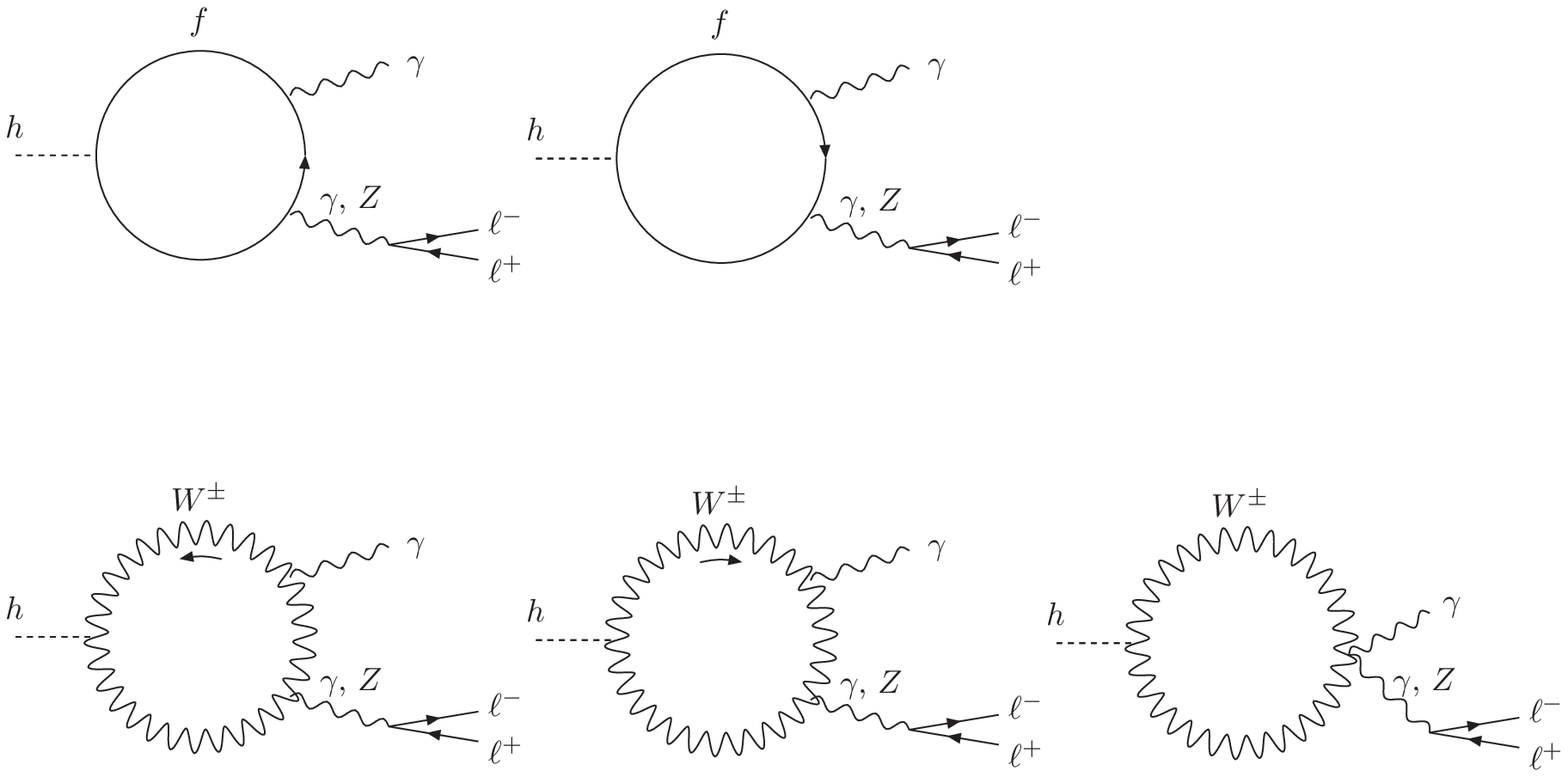}
\end{center}
\caption{Diagrams of the process $h \to \gamma \ell^+ \ell^-$: the tree-level diagrams are shown on the top, 
one-loop diagrams -- on the bottom. Fermions $f$ are denoted by solid lines, gauge bosons 
$W^\pm, \, Z, \, \gamma$ -- by wavy lines and $h$ boson -- by dashed lines }
\label{fig:diagrams}
\end{figure}

The decay amplitude is
\begin{equation}\label{eq:004}
{\cal M} \, = \, {\cal M}_{tree} \, + \, {\cal M}_{loop}\, ,
\end{equation}
where the tree-level amplitude (see Fig.~\ref{fig:diagrams}) is
\begin{eqnarray}
{\cal M}_{tree} &=& c_0 \, \epsilon_\mu^* (k) \, \bar{u}(q_-) (a_{\ell}+i \,b_{\ell} \gamma_5 ) \nonumber  \\
& \times &  \Bigl( \frac{2q_+^\mu + \not{\!k} \gamma^\mu}{2 k
\cdot q_+ } - \frac{2q_-^\mu + \gamma^\mu \not{\!k}}{2 k \cdot q_-
} \Bigr) \, v(q_+)  \, , \label{eq:005}
\end{eqnarray}
with
\begin{equation}
c_0 = e m_{\ell} Q_{\ell} \, ({\sqrt{2} G_F})^{1/2}\, ,
\label{eq:c_0}
\end{equation}
$e = \sqrt{4 \pi \alpha_{G_F}}$ is the positron electric charge,
$Q_{\ell} = -1$ (lepton charge in units of $e$) and $m_{\ell}$ is
the lepton mass. The electromagnetic coupling constant in the
$G_F$ scheme~\cite{Heinemeyer:2013} is \ $\alpha_{G_F} =
\sqrt{2}G_F m_W^2(1-m_W^2 /m_Z^2)/ \pi$, where $m_W \, (m_Z)$ is
the mass of the $W$ ($Z$) boson.

The one-loop contributions to the $h \to \gamma \, \gamma^* / Z^* \to \gamma
\ell^+ \ell^-$ decay (see Fig.~\ref{fig:diagrams}) can be written in the form
\begin{eqnarray}
 {\cal M}_{loop} &=& \epsilon_\mu^* (k) \, [ \, (q^\mu k^\nu -  g^{\mu \nu} k\cdot q)   \nonumber \\
& \times &  \bar{u}(q_-) ( c_1 \gamma_\nu + c_2 \gamma_\nu
\gamma_5) v(q_+)  \nonumber
\\
&-& \epsilon^{\mu \nu \alpha \beta} k_\alpha q_\beta \,
\bar{u}(q_-) ( c_3 \gamma_\nu + c_4 \gamma_\nu \gamma_5 ) v(q_+)
\,  ] \,  \label{eq:006}
\end{eqnarray}
with the coefficients $c_1, \ldots , c_4 $ which have been
obtained  in~\cite{Korchin:2014} and are collected for convenience
in Appendix~\ref{sec:appendix}. Besides,  $\epsilon_{0123} = +1$.

In the present work we do not include the box-type loop
contributions to the process $h \to \gamma \ell^+ \ell^- $. The
contribution from these diagrams in the SM is very
small~\cite{Abbasabadi:1995,Abbasabadi:1997}. In addition, there are
other mechanisms, $h \to \gamma V \to \gamma \ell^+ \ell^-$, where
$V$ is intermediate vector resonance decaying to the $\ell^+
\ell^-$ pair, which can contribute to the decay $h \to \gamma
\ell^+ \ell^-$. Particularly, the production of charmonium $J/ \psi
\, (c \bar{c})$ and bottomonium $\Upsilon (1S) \, (b \bar{b})$ is
interesting for studying the $h q \bar{q}$ \ interaction (see, {\it e.g.}, \cite{Bodwin:2013,Kagan:2014,Gao:2014,London:2014}).
However, these processes lie beyond the scope of the present work.

Calculating the amplitude (\ref{eq:004}) squared and summed over
the polarizations of leptons and photon we obtain in the model
(\ref{eq:001})
\begin{eqnarray}\label{eq:007}
\sum_{\rm
pol}|{\cal M}|^2  &=&  c_0^2 \, \bigl[  |a_{\ell}|^2 \, A + |b_{\ell}|^2 \, \widetilde{A}  \bigr] \nonumber \\
&& + 2 \, c_0 \, \bigl[{\rm Re} (c_1\,a_{\ell}^*) \, B + {\rm Im} (c_2\,b_{\ell}^*) \, \widetilde{B} \, \nonumber \\
&& + {\rm Im} (c_4\,a_{\ell}^*) \, C + {\rm Re}(c_3\,b_{\ell}^*) \, \widetilde{C} \bigr] \nonumber \\
&& + \bigl( |c_1|^2 + |c_3|^2  \bigr) \, D  + \bigl( |c_2|^2 + |c_4|^2  \bigr) \, E  \nonumber \\
&& + 2 \, {\rm Im} \bigl(c_1 c_4^* + c_2 c_3^*\bigr) \, F \, ,
\end{eqnarray}
where $A$, $\widetilde{A}$, $B$, $\widetilde{B}$, $C$,
$\widetilde{C}$, $D$, $E$, $F$ have the same form as in~\cite{Korchin:2014} 
and are given in Appendix~\ref{sec:appendix}.

The forward-backward (FB) asymmetry is defined (see, {\it e.g.},
\cite{Korchin:2014})
\begin{equation}\label{eq:008}
A_{\rm FB} (q^2) =\Big( {\frac{{\rm d} \Gamma_F }{{\rm d} q^2} \,
- \, \frac{{\rm d} \Gamma_B }{{\rm d} q^2}} \Big)  \Big/ {\Big(
\frac{{\rm d} \Gamma_F }{{\rm d} q^2} \,+ \, \frac{{\rm d}
\Gamma_B }{{\rm d} q^2} \Big) }\,,
\end{equation}
where
\begin{eqnarray}
\frac{{\rm d} \Gamma_F }{{\rm d} q^2} & \equiv &
\int_0^1 \frac{{\rm d} \Gamma}{{\rm d} q^2 \, {\rm d} \cos\theta}\, {\rm d} \cos\theta \,, \nonumber  \\
\frac{{\rm d} \Gamma_B }{{\rm d} q^2} &\equiv & \int_{-1}^0 \frac{
{\rm d} \Gamma}{ {\rm d} q^2 \, {\rm d} \cos\theta}\, {\rm
d}\cos\theta \, , \, \label{eq:009}
\end{eqnarray}
and also the $q^2$ integrated FB asymmetry reads 
\begin{equation}
\langle A_{\rm FB}\rangle= \Big\langle{\frac{{\rm d} \Gamma_F }{{\rm d}
q^2} \, - \, \frac{{\rm d} \Gamma_B }{{\rm d}
q^2}} \Big\rangle \Big/ \Big\langle{\frac{{\rm d} \Gamma_F }{{\rm d} q^2} \, + \,
\frac{{\rm d} \Gamma_B }{{\rm d} q^2}} \Big\rangle\,,\label{eq:010}
\end{equation}
with the notation
\begin{equation}
\langle J\rangle\equiv \int_{q^2_{\rm min}}^{q^2_{\rm max}}d q^2
J(q^2)\,\label{eq:011}
\end{equation}
for the integration limits $q^2_{\rm min}\geq 4 m_\ell^2$ and $q^2_{\rm max}\leq m_h^2$.

In Eqs.~(\ref{eq:007}), only the coefficients $\widetilde{B}$, $C$
and  $F$ are linear in $\cos \theta$, therefore, as can be seen
from (\ref{eq:003}), (\ref{eq:007}) and (\ref{eq:008}), the
numerator of the FB asymmetry (\ref{eq:008}) is determined by 
the imaginary combination of the terms
$c_2\,b_\ell^*+c_4\,a_\ell^*$ and $c_1\,c_4^*+c_2\,c_3^*$:
\begin{eqnarray}
&& \frac{{\rm d} \Gamma_F }{{\rm d} q^2} - \frac{{\rm d} \Gamma_B
}{{\rm d} q^2} =
-\frac{2  (m_h^2-q^2)^2 }{(8 \pi)^3 \, m_h^3} \nonumber \\
&& \times \Big[ 8 m_\ell\, c_0  \, {\rm Im}(c_2\,b_\ell^*+c_4\,a_\ell^*) \ln \Big( \frac{q^2}{4m_\ell^2} \Big)  \nonumber \\
&& + \, {\rm Im} \big(c_1 c_4^* + c_2 c_3^*\big) \, (q^2  - 4
m_\ell^2) \, (m_h^2-q^2)  \Big]\, . \label{eq:delta_Gamma}
\end{eqnarray}
In framework of the SM the differential decay width
(\ref{eq:003}), (\ref{eq:007}) takes a simpler form. In this case,
$a_{\ell}=1$, $b_{\ell}=0$ and $c_{3, {\rm SM}} = c_{4, {\rm
SM}} = 0$. Thus the FB asymmetry vanishes,
\begin{equation}
A_{\rm FB } (q^2)_{\rm SM} = 0\, .  \label{eq:A_FB_SM}
\end{equation}
Therefore, nonzero values of this asymmetry can arise only in certain models of new physics.

\section{ \label{sec:results}  Results of calculations and discussion}

First, we discuss the choice of parameters $a_f$ and $b_f$, that
determine the interaction of the Higgs boson with fermions
(\ref{eq:001}). In the leading order, the rate of the $h$-boson decay to fermions,
except the top quarks, has the form
\begin{equation}
\Gamma (h \to f \bar{f})\, = \, \frac{N_f G_F}{4 \sqrt{2} \pi} \,
m_f^2 \, m_h \, \beta_f \bigl( |a_f|^2 \beta_f^2 \, + \, |b_f|^2
\bigr) \,,
\label{eq:width_hff}
\end{equation}
where $\beta_f = \sqrt{1- 4m_f^2/m_h^2} $ is the fermion velocity
in the rest frame of $h$,  \ $N_f=1 \, (3)$ for leptons (quarks).
Apparently, $\beta_f$ is equal to one with a good accuracy. We
assume further that the $h \to f \bar{f}$ decay rate in the model
(\ref{eq:001}) is the same as in the SM,  {\it i.e.},
\begin{equation} |a_f|^2 + |b_f|^2 = 1 \,. \label{eq:s_f+p_f}
\end{equation}
In this case, to search for effects of new physics in the decay $h
\to f \bar{f}$, it will be necessary to measure the polarization
characteristics of the fermions, which clearly complicates the
identification of particle $h$ with the Higgs boson of the SM.

Let us calculate the predictions of the model (\ref{eq:001}) with
the constraint  $|a_f|^2 + |b_f|^2 = 1$ for the decay $h \to
\gamma \ell^+ \ell^-$ and check how much these predictions differ
from the SM.
In the calculations we use the parameters of NP
presented  in Table~\ref{tab:new physics}.

\vspace{0.1cm}

\begin{table}[tbh]
\caption{Parameters of the $h f \bar{f}$ interaction in the SM,
and in several models of NP. Here, $\ell$ denote leptons, $q$
denote quarks of all flavors, except the top quark, for which
parameters are shown separately }
\begin{center}
\begin{tabular}{c |c  c c c }
\hline \hline
$h f \bar{f}$ couplings   & SM     & NP1  &  NP2   & NP3   \\
\hline
$a_{\ell}$  &  1 & $\frac{1}{\sqrt{2}}$  & $\frac{1}{\sqrt{2}}$  & $\frac{1}{\sqrt{2}}$ \\
$b_{\ell}$  & 0 & $\frac{1}{\sqrt{2}}$  & $\frac{i}{\sqrt{2}}$  & $\frac{i}{\sqrt{2}}$  \\
$a_{q}$  & 1  & $\frac{1}{\sqrt{2}}$  & $\frac{1}{\sqrt{2}}$  & $1 $   \\
$b_{q}$  & 0 & $\frac{1}{\sqrt{2}}$  & $\frac{i}{\sqrt{2}}$  & $0 $  \\
$a_{t}$  & 1  & $1.2$  & $1.2$  & $1 $   \\
$b_{t}$  & 0 & $0.37$  & $0.37  i$  & $0 $  \\
\hline  \hline
\end{tabular}
\end{center}
\label{tab:new physics}
\end{table}

The new physics model NP1 is described by the parameters
$a_f=1/\sqrt{2},\; b_f=1/\sqrt{2}$ for all fermions, except the
top quark for which the couplings are taken
from~\cite{Kobakhidze:2016}. In the NP1 model, the interaction
(\ref{eq:001}) is Hermitian.
In the model NP2, $a_f=1/\sqrt{2},\; b_f=i/\sqrt{2}$, while for the
top quark we choose $a_t=1.20,\; b_t=0.37 i$, so that the
interaction (\ref{eq:001}) becomes non-Hermitian. Finally, in the
model NP3, the interaction for the leptons is non-Hermitian, while
for all  quarks the interaction (\ref{eq:001}) is taken as in the
SM.

The numerical values of other parameters of the SM are taken
from~\cite{PDG:2016}, in particular, the masses of the $W^\pm$ and
$Z$ bosons, the decay widths and couplings for $Z f \bar{f}$
interaction. The quark masses are chosen as
in~\cite{Heinemeyer:2013,Dittmaier:2011}, and $\sin^2 \theta_W =
1- m_W^2/m^2_Z$.

\begin{figure}[tbh]
\begin{center}
\includegraphics[width=0.45\textwidth]{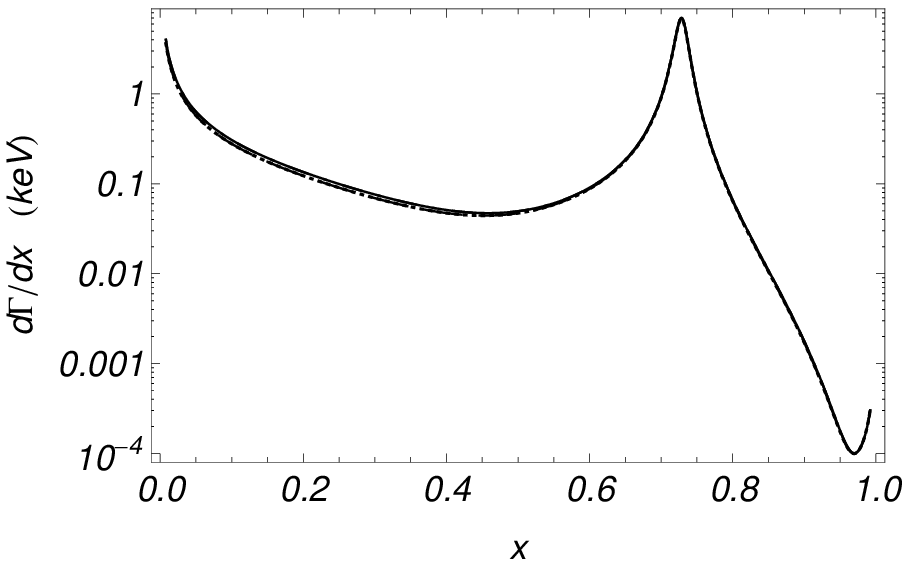}
\includegraphics[width=0.45\textwidth]{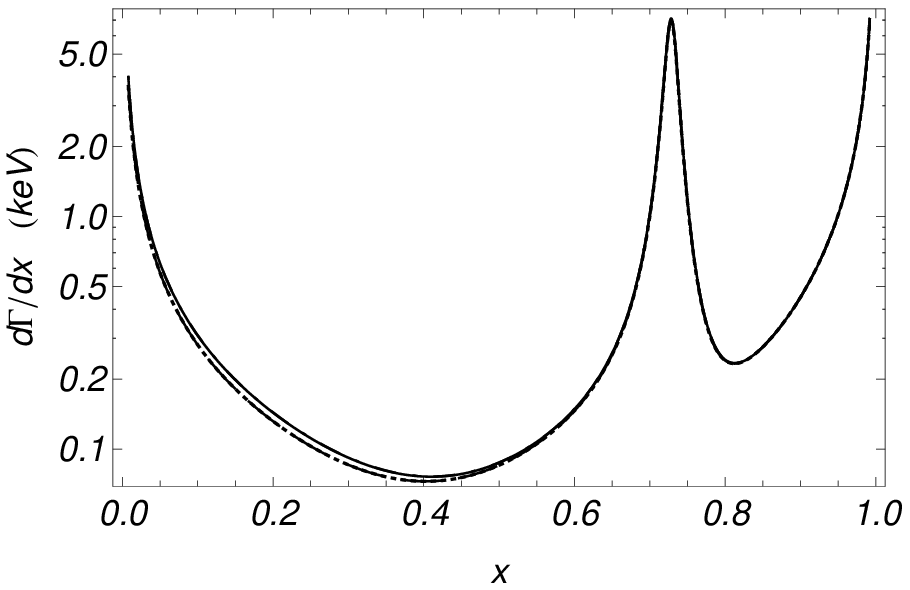}
\includegraphics[width=0.45\textwidth]{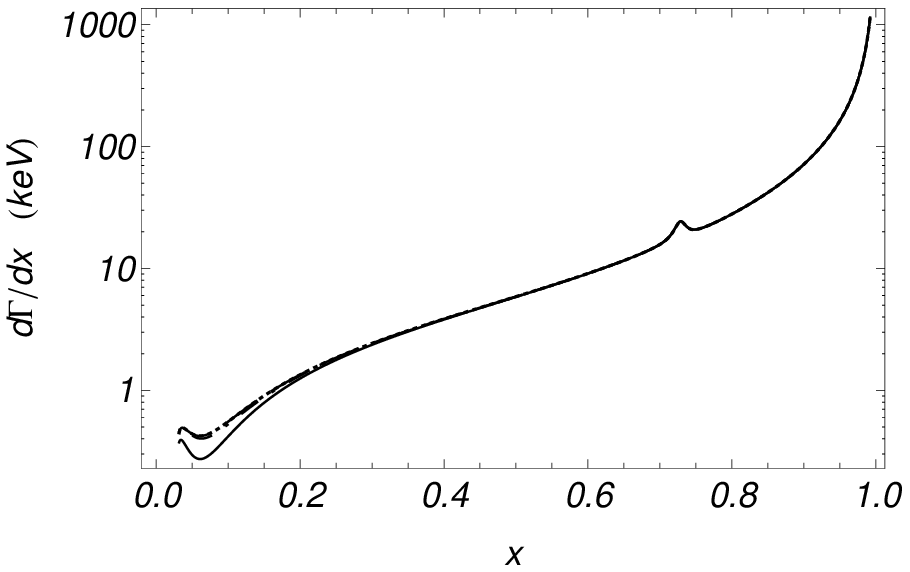}
\end{center}
\caption{Differential decay width of $h \to \gamma \ell^+ \ell^-$
for various lepton pairs as a function of $x$, where $x\equiv
\sqrt{q^2}/m_h$. The top part is drawn for $e^+e^-$ pair, the middle part -- $\mu^+ \mu^-$ pair, 
and the bottom part -- $\tau^+ \tau^-$ pair.  The solid lines correspond to the SM, dotted lines -- model NP1,  
dashed lines  -- model NP2, dash-dotted lines -- model NP3 }
\label{fig:widths}
\end{figure}

In Fig.~\ref{fig:widths} we present the differential width of the
$h \to \ell^+ \ell^- \gamma$ decay for various leptons  $l = (e,
\, \mu, \, \tau)$ calculated in the SM and in the model
(\ref{eq:001}) with the couplings in Table~\ref{tab:new physics}.
The minimal photon energy in the rest frame of $h$ boson is chosen
$E_\gamma^{\rm min} = 1$ GeV to cut off the infrared divergence at
$E_\gamma \to 0$. This leads to maximal value of the dilepton
invariant mass $q_{\rm max} = (m_h^2 - 2 m_h E_\gamma^{\rm min})^{1/2}
\approx m_h - E_\gamma^{\rm min}=124$ GeV for $m_h=125.09$ GeV.

As one can see from Fig.~\ref{fig:widths}, there are deviations
from the SM predictions with the chosen parameters $a_f,\, b_f$.
In Table~\ref{tab:widths}, we also show the decay widths integrated
over the invariant mass within the limits $[q_{\rm min}, \, q_{\rm max}]$.

\vspace{0.1cm}

\begin{table}[tbh]
\caption{The width of the decay $\Gamma (h \to \gamma \ell^+
\ell^-)$ (in keV) for various lepton pairs in the invariant mass
limits $[q_{\rm min}, \, q_{\rm max}]$ (in GeV)}
\begin{center}
\begin{tabular}{c c c c c c c}
\hline  \hline
$\ell^+ \ell^-$  & $q_{\rm min}$  & $q_{\rm max}$  &  SM &  NP1 & NP2 & NP3 \\
\hline
$e^+ e^-$   & $1.0$  & $124.0$ & $0.34$ & $0.32$ & $0.32$ & $0.34$ \\
$ $         & $1.0$  & $30.0$  & $0.11$ & $0.10$ & $0.10$ & $0.11$ \\
$ $         & $37.5$ & $75.0$  & $0.02$ & $0.02$ & $0.02$ & $0.02$ \\
\hline
$\mu^+ \mu^-$ & $1.0$ & $124.0$ & $0.53$ & $0.52$ & $0.52$ & $0.53$  \\
$ $           & $1.0$ & $30.0$  & $0.11$ & $0.10$ & $0.10$ & $0.11$  \\
$ $           & $37.5$ & $75.0$ & $0.03$ & $0.03$ & $0.03$ & $0.03$  \\
\hline
$\tau^+ \tau^- $ & $4.0$ & $124.0$ & $31.0$ & $31.1$ & $31.1$ & $31.1$ \\
$ $              & $12.5$ & $75.0$ & $1.77$  & $1.80$  & $1.80$  & $1.79$  \\
\hline \hline
\end{tabular}
\end{center}
\label{tab:widths}
\end{table}

In the interval of invariant masses from $1.0$ GeV to $124.0$ GeV,
the effect of new physics does not exceed 5\% in the decays $h \to
\gamma e^+ e^-$ and $h \to \gamma \mu^+ \mu^-$.  However, at small
invariant masses below 30 GeV, this effect reaches 10\%, although
the decay rate in this interval is very small compared, for
example, with the rate of the Higgs boson decay to two photons
 $\Gamma (h \to \gamma \gamma)=9.28$ keV~\cite{Heinemeyer:2013}.

\begin{figure}[tbh]
\begin{center}
\includegraphics[width=0.45\textwidth]{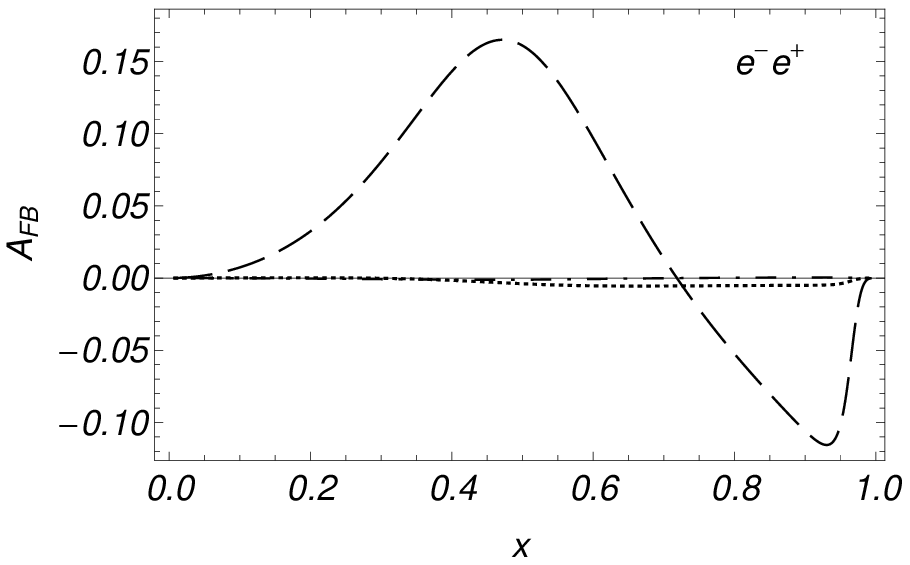}

\includegraphics[width=0.45\textwidth]{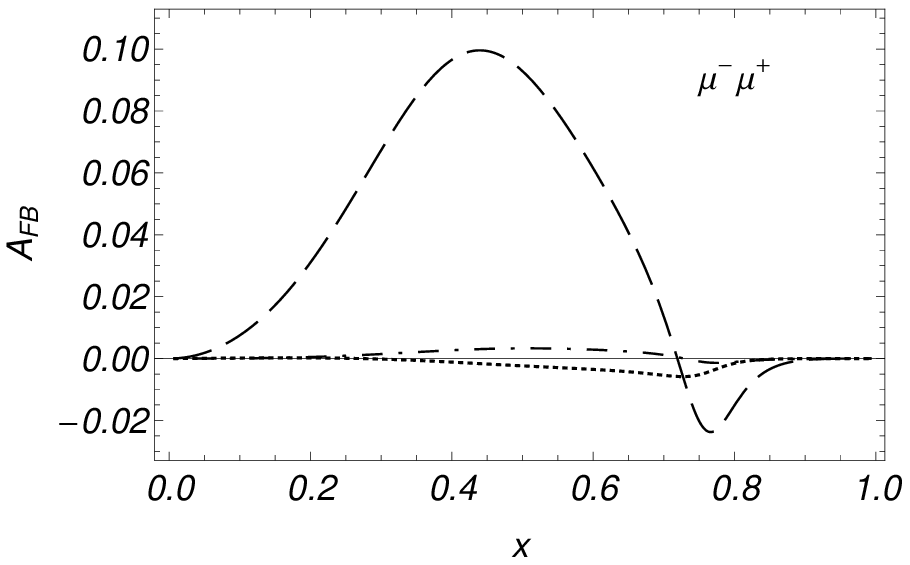}

\includegraphics[width=0.45\textwidth]{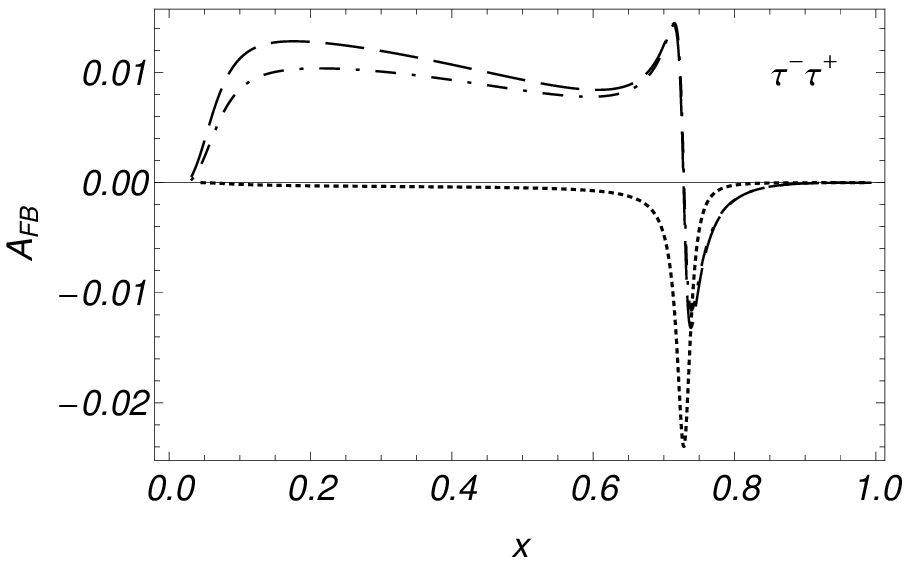}

\end{center}
\caption{Forward-backward asymmetry in the decay $h \to \gamma
\ell^+ \ell^-$ for various lepton pairs as a function of $x$,
where $x\equiv \sqrt{q^2}/m_h$. The dotted lines correspond to
NP1, dashed lines  -- NP2, dash-dotted lines -- NP3}
\label{fig:asymmetries}
\end{figure}

In Fig.~\ref{fig:asymmetries}, we show the forward-backward
asymmetry (\ref{eq:008}).  As was previously mentioned, the FB
asymmetry takes zero value in the SM, and non-zero values can
arise only in models beyond the SM. Of course, not all models of NP
lead to non-zero FB asymmetry.

In Table~\ref{tab:asymmetries} we also present the integrated FB asymmetry (\ref{eq:010}).


\begin{table}[tbh]
\caption{Forward-backward asymmetry in the decay $\Gamma (h \to
\gamma \ell^+ \ell^-)$ in \% for various leptons in the limits of invariant dilepton mass from
$q_{\rm min}$ to $q_{\rm max}$ (in GeV)}
\begin{center}
\begin{tabular}{c c c c c c }
\hline  \hline
$\ell^+ \ell^-$  & $q_{\rm min}$  & $q_{\rm max}$  &  NP1 & NP2 & NP3 \\
\hline
$e^+ e^-$   & $1.0$  & $124.0$  & $-0.36$ & $0.78$ & $-0.01$ \\
$ $         & $1.0$  & $30.0$   & $0.01$  & $0.61$ & $-0.01$ \\
$ $         & $37.5$ & $75.0$   & $-0.28$ & $13.1$ & $-0.1$ \\
\hline
$\mu^+ \mu^-$ & $1.0$ & $124.0$ & $-0.24$ & $0.52$ & $0.03$  \\
$ $           & $1.0$ & $30.0$  & $0.01$  & $0.62$ & $0.01$  \\
$ $           & $37.5$ & $75.0$ & $-0.19$ & $8.5$  & $0.28$  \\
\hline
$\tau^+ \tau^- $ & $4.0$ & $124.0$ & $-0.06$ & $0.08$ & $0.07$ \\
$ $              & $12.5$ & $75.0$ & $-0.05$ & $1.0$  & $0.88$  \\
\hline  \hline
\end{tabular}
\end{center}
\label{tab:asymmetries}
\end{table}

As it is seen from Fig.~\ref{fig:asymmetries}, the FB asymmetry
for the $e^+e^-$ and $\mu^+\mu^-$ pairs for the real parameters
$a_f$ and $b_f$ is very small, of the order of 1\%. Only for the
pair of the $\tau$-leptons, this asymmetry reaches 2.5\% at the
invariant mass near the $Z$-boson mass.

The FB asymmetry for the $e^+e^-$ and $\mu^+ \mu^-$ pairs
increases considerably for the non-Hermitian $h f \bar{f}$
interaction in the model NP2. Namely, the FB asymmetry can
reach 15\% for the electron-positron pair and 10\% for the
muon-antimuon pair. In the model NP3, in which only interaction $h
\ell^+ \ell^-$ is non-Hermitian, the FB asymmetry is still very
small. Thus, the most important contribution comes from
non-Hermiticity of the Higgs boson interaction with the quarks,
mainly with the top quark.  This $h t \bar{t}$ interaction enters
in the loop diagrams in Fig.~\ref{fig:diagrams}.

As for the $\tau^+ \tau^-$ pair, the behavior is different from
the case of the light leptons. In the process $h \to \tau^+ \tau^-
\gamma$, the dominant contribution comes from the tree-level
diagram in Fig.~\ref{fig:diagrams}, and therefore the structure of
the interaction of Higgs boson with tau lepton is crucial. This
explains the observed tendency in Fig.~\ref{fig:asymmetries}, in
which the models NP2 and NP3 give the close results, though the
absolute value of the FB asymmetry does not exceed 1.5\%.

It is seen from Fig.~\ref{fig:asymmetries} that the FB asymmetries
change sign as functions of the variable $x = \sqrt{q^2}/m_h$.
Therefore, integration of the asymmetries over the whole interval
of invariant masses gives small values. This is demonstrated in
Table~\ref{tab:asymmetries}: the obtained values for the $e^+ e^-$
and $\mu^+ \mu^- $ pairs are less then 1\%, and for $\tau^+
\tau^-$ pair -- less than 0.1\%. However, choosing the suitable
intervals of integration increases the corresponding values to
13.1\% for the $e^+ e^-$ pair, and 8.5\% for the $\mu^+ \mu^- $
pair. For the $\tau^+ \tau^-$ pair, the integrated FB asymmetry
does not exceed 1\%.


\section{ \label{sec:conclusions} Conclusions}

We studied the decay of the Higgs boson to a photon and a
lepton-antilepton pair, $h \to \gamma \ell^+ \ell^-$,
where $\ell = (e,\, \mu, \, \tau)$. The differential decay width
and lepton forward-backward asymmetry are calculated as functions
of the dilepton invariant mass.

These observables are calculated in the Standard model and in the
model in which the Higgs boson interaction with the fermions
consists of scalar and pseudoscalar terms, which imply the $\mathcal{CP}$
violation. Moreover, we assume a possible non-Hermiticity of this
interaction. The tree-level amplitudes and the one-loop $h \to
\gamma Z^* \to \gamma \ell^+ \ell^-$ and $h \to \gamma \gamma^*
\to \gamma \ell^+ \ell^-$ diagrams are included. The main emphasis
is put on studying effects of possible non-Hermiticity of the $h f
\bar{f} $ interaction on the observables.

The observables are calculated with the model parameters $a_f, \,
b_f$ chosen in such a way that the $h \to f \bar{f}$  decay rate
(where $f = (\ell, \, q)$) coincides with the rate in the SM. For
the couplings with the top quark, $a_t, \, b_t$, we choose the
values from Ref.~\cite{Kobakhidze:2016} which are constrained from
all available data.

The calculations show that the differential decay width for $h \to
\gamma e^+ e^- $ and $h \to \gamma \mu^+ \mu^-$ is not very
sensitive to effects of NP. Only at small values of the dilepton
invariant mass below 30~GeV, the corrections to the SM prediction
reach 10\%, though the decay rate in this interval is rather
small, about 0.1~keV.

The lepton forward-backward asymmetry $A_{\rm FB}$ is more sensitive
to effects of NP, because this observable vanishes
identically in the SM. However, for real parameters of the $h f \bar{f}$
interaction, $A_{\rm FB}$ is small, of the order of 1\% for the $e^+
e^-$ and $\mu^+ \mu^- $ pairs, and 2.5\% for the $\tau^+ \tau^- $
pair. This asymmetry becomes sizable for the complex parameters $a_f,
\, b_f$, {\it i.e.,} for the non-Hermitian interaction. In particular,
for the $e^+ e^-$ and $\mu^+ \mu^- $ pairs, $A_{\rm FB}$ rises to 15\%
for the $e^+ e^-$ pair and to 10\% for the $\mu^+ \mu^- $ pair. The
main contribution to $A_{\rm FB}$ comes from the non-Hermitian interaction
of the Higgs boson with the top quark in  the loop diagrams.

At the same time, for the $\tau^+ \tau^- $ pair, the tree-level
diagrams are dominant, and thus the asymmetry depends on the Higgs
interaction with the tau leptons. It turns out that effect of a
non-Hermiticity in $A_{\rm FB}$ is of the order of 1\%, which may be
difficult for experimental studies.

Our consideration of the decays $h \to \gamma \ell^+ \ell^-$
demonstrates that a possible non-Hermiticity of the $h t \bar{t}$ \
interaction has big impact on the forward-backward asymmetry for
the light leptons. The non-Hermiticity of the $h \ell^+ \ell^-$
interaction does not show up in this observable.
In summary, the forward-backward
asymmetry in the $h \to \gamma e^+ e^-$ and $h \to \gamma \mu^+
\mu^-$ decays is the informative and important observable for
experimental studies at the LHC in the search for effects of new
physics.

\begin{acknowledgements}
This research was partially supported by the Ministry of Education
and Science of Ukraine (projects no. 0115U000473 and 0117U004866)
and the National Academy of Sciences of Ukraine (project TsO-1-4/2016).
\end{acknowledgements}

\appendix

\section{Definition of coefficients $c_1, \ldots , c_4$ and $A$, $\widetilde{A}$, $B$, $\widetilde{B}$, $C$,
$\widetilde{C}$, $D$, $E$, $F$.  }
\label{sec:appendix}

In this Appendix we present the coefficients $c_1, \ldots , c_4$
in Eq.~(\ref{eq:006}), which are determined from the loop diagrams
in Fig.~{\ref{fig:diagrams}}. They read~\cite{Korchin:2014}:
\begin{eqnarray}
c_1 &=&  \frac{1}{2} \,  \frac{g_{V,  \ell}}{q^2 - m_Z^2 + i m_Z \Gamma_Z} \,
\Pi_{ Z} + \frac{Q_{\ell}}{q^2} \,  \Pi_{\gamma } \, , \label{eq:c_1} \\
c_2 &=&  - \frac{1}{2} \, \frac{g_{A,  \ell}}{q^2 - m_Z^2 + i m_Z \Gamma_Z} \, \Pi_{ Z} \, , \nonumber \label{eq:c_2} \\
c_3 &=&  \frac{1}{2} \,  \frac{g_{V,  \ell}}{q^2 - m_Z^2 + i m_Z \Gamma_Z} \,  \widetilde\Pi_{Z} + \frac{Q_{\ell}}{q^2} \, \widetilde\Pi_{\gamma } \, , \nonumber \label{eq:c_3}  \\
c_4 &=&   - \frac{1}{2}  \, \frac{ g_{A,  \ell}}{q^2 - m_Z^2 + i m_Z \Gamma_Z} \, \widetilde\Pi_{ Z} \, \nonumber \label{eq:c_4} ,
\end{eqnarray}
with
\begin{eqnarray}
\Pi_{ Z} &=&   \frac{e g^3}{16 \pi^2 m_W } \,
\bigl[ a_f \, \frac{2 g_{V,  f}}{c_W^2} \, N_f \, Q_f \, A_f (\lambda_f^\prime, \lambda_f)   \nonumber \\
&&  +  A_W (\lambda_W^\prime, \lambda_W) \bigr] \, , \label{eq:Pi_Z} \\
\Pi_{\gamma} &=&   \frac{e^3 g}{16 \pi^2 m_W} \,
\bigl[ a_f \, 4 Q_f^2 \, N_f  \, A_f (\lambda_f^\prime, \lambda_f)  \nonumber \\
&& + A_W (\lambda_W^\prime, \lambda_W) \bigr] \, ,   \label{eq:Pi_gamma}
\end{eqnarray}
\begin{eqnarray}
\widetilde\Pi_{ Z} &=&   \frac{e g^3}{16 \pi^2m_W  } \, b_f \, \frac{2 g_{V,f}}{c_W^2} \, N_f \, Q_f \,  I_2 (\lambda_f^\prime, \lambda_f)  ,  \label{eq:tilde-Pi_Z}  \\
\widetilde\Pi_{ \gamma} &=&   \frac{e^3 g}{16 \pi^2 m_W} \, b_f \, 4 Q_f^2 \, N_f \, I_2 (\lambda_f^\prime, \lambda_f) \, . \label{eq:tilde-Pi_gamma}
\end{eqnarray}
Here, $\Gamma_Z$ is the total decay width of the
$Z$ boson, $c_W \equiv \cos \theta_W$, where $\theta_W$ is the
weak angle, $g = 2 m_W (\sqrt{2}G_F)^{1/2}$, $Q_f$ is the charge of the fermion in units
of $e$, \ $g_{V,  f} = t_{3L, f} - 2Q_f s_W^2$ \ ($g_{A,  f} = t_{3L, f}$) is the $Z f \bar{f}$ vector coupling (axial-vector one), where $t_{3L, f}$ is
projection of the weak isospin.
The sum over all leptons and quarks in
(\ref{eq:Pi_Z})-(\ref{eq:tilde-Pi_gamma}) is implied.

The loop integrals for fermions, $A_f (\lambda_f^\prime,
\lambda_f)$,  and $W$ bosons,  $A_W (\lambda_W^\prime,
\lambda_W)$, are expressed via the loop functions $I_{1}
(\lambda^\prime, \, \lambda)$ and  $I_{ 2} (\lambda^\prime, \,
\lambda)$ introduced in~\cite{Spira:1998} and given explicitly in~\cite{Korchin:2014}.
The arguments of these loop functions are
\begin{equation}
 \lambda_{f, \, W} \equiv {4m_{f, \, W}^2}/{q^2 }\, , \qquad \lambda^{\prime}_{f, \, W} \equiv \lambda_{f, \, W} |_{q^2 = m_h^2} \,.
\label{eq:tau_lambda}
\end{equation}

Further, the coefficients in Eq.~(\ref{eq:007}) are defined as follows:
\begin{eqnarray}
A&=& \frac{16}{(1- \beta_{\ell}^2 z^2 )^2 (m_h^2-q^2)^2}
[\, (m_h^4 + q^4  \nonumber \\
&& - 8 m_{\ell}^2 q^2)(1-\beta_{\ell}^2 z^2) +32 m_{\ell}^4 - 8 m_h^2 m_{\ell}^2 \, ]\, , \label{eq:ap_008} \\
\widetilde{A} &=& \frac{16}{(1- \beta_{\ell}^2 z^2 )^2 (m_h^2-q^2)^2}
[\, (m_h^4+q^4) \nonumber   \\
&& \times \, (1-\beta_{\ell}^2 z^2)
 -  8 m_h^2 m_{\ell}^2\, ]\, , \nonumber \label{eq:ap_009}
\end{eqnarray}
\begin{eqnarray}
B &=& - \frac{8 m_{\ell} }{(1- \beta_{\ell}^2 z^2 )}  [\, m_h^2 -q^2 +q^2 \beta_{\ell}^2 (1-z^2 )\, ] \, ,  \nonumber \label{eq:ap_010} \\
\widetilde{B} &=&  - \frac{8 m_{\ell}  }{(1- \beta_{\ell}^2 z^2 )} \,  (m_h^2-q^2) \, \beta_{\ell} \, z \, , \nonumber \label{eq:ap_011}\\
C & =& - \frac{8 m_{\ell} }{(1- \beta_{\ell}^2 z^2 )} \, (m_h^2-q^2) \, \beta_{\ell} \, z \, , \nonumber \label{eq:ap_012} \\
\widetilde{C} &=& \frac{8 m_{\ell} }{(1- \beta_{\ell}^2 z^2 )} (m_h^2-q^2)\, , \nonumber \label{eq:ap_013} \\
D &=& \frac{1}{2} (m_h^2-q^2)^2 \, [\, q^2 (1 + \beta_{\ell}^2 z^2  ) +4 m_{\ell}^2 \,]\, , \nonumber \label{eq:ap_014} \\
E &=&  \frac{1}{2} (m_h^2-q^2)^2 \, q^2 \, \beta_{\ell}^2 \, (1 + z^2 )\, , \nonumber \label{eq:015} \\
F &=& - (m_h^2-q^2)^2 \, q^2 \, \beta_{\ell} \,  z \, \nonumber \label{eq:ap_016}
\end{eqnarray}
with the notation $z \equiv \cos \, \theta$.


\end{document}